\documentstyle[aps,preprint,pra]{revtex}
\tighten
\begin{document}
%\twocolumn[\hsize\textwidth\columnwidth\hsize\csname @twocolumnfalse\endcsname
\draft
\title{The classical communication cost of entanglement manipulation: Is 
entanglement an inter-convertible resource?}
\author{Hoi-Kwong
 Lo \thanks{e-mail: hkl@hplb.hpl.hp.com}}
\address{Hewlett-Packard Labs, Filton Road, Stoke Gifford, Bristol
 BS34 8QZ, United Kingdom}
\author{and}
\author{Sandu Popescu \thanks{email: sp230@newton.cam.ac.uk}}
\address{Isaac Newton Institute for Mathematical Sciences, Cambridge, CB3 0EH, 
UK and
BRIMS, Hewlett-Packard Labs, Filton Road, Stoke Gifford, Bristol
BS34 8QZ, United Kingdom}

\date{\today}
\maketitle
\begin{abstract}

Entanglement bits or ``ebits'' have been proposed as a quantitative measure of
a fundamental resource in quantum information processing.  For such an
interpretation to be valid, it is important to show that the same number of
ebits in different forms or concentrations are inter-convertible in the
asymptotic limit.  Here we draw attention to a very important but hitherto
unnoticed aspect of entanglement manipulation --- the classical communication
cost.  We construct an explicit procedure which demonstrates
that for bi-partite pure states, in the asymptotic limit, entanglement can
be concentrated or diluted with vanishing classical communication cost.
Entanglement of bi-partite pure states is thus established as a truly 
inter-convertible resource.

\end{abstract}

\medskip
\pacs{\noindent
 \begin{minipage}[t]{5in}
%  PACS numbers: 03.65.Bz, 89.80.+h,  \vspace{1ex}
% \\
 \end{minipage}
}

%\pacs{PACS numbers: 03.65.Bz, 89.70.+c, 89.80.+h}
%]
%\narrowtext
%\footnotetext[1]{e-mail: hfchau@hkusua.hku.hk}
%\footnotetext[2]{e-mail: hkl@hplb.hpl.hp.com}

During the last couple of years the study of quantum non-locality (entanglement) 
has undergone a substantial transformation. It has become clear that 
entanglement is
a most important aspect of quantum mechanics, which plays a 
fundamental role in quantum information
processing (including teleportation~\cite{tele},
dense coding~\cite{dense}, and
communication complexity~\cite{complexity}).
It is now customary to regard entanglement as a fungible resource, 
i.e., a resource which can be transformed from one form to another, can be 
created, stored or consumed for accomplishing useful tasks. It is however the 
aim 
of this paper to draw attention to an important and hitherto ignored aspect 
of entanglement manipulation which has to be clarified before one can regard 
entanglement as a completely fungible property. The problem is
the {\it 
classical information cost} of entanglement manipulation.

Consider the most famous use of entanglement, namely teleportation.  As Bennett 
{\it et al.}~\cite{tele} have shown, entanglement can be
used to communicate unknown 
quantum states from one place to another; this task can be achieved even though 
neither the transmitter nor the receiver are able to find out the 
state to be transmitted.
 
The basic equation of teleportation is

\begin{equation}
1~{\rm singlet}~teleports~1~{\rm qubit}.
\end{equation}

Equation (1) already contains a large degree of abstraction.  In the original
description of teleportation it was shown how a singlet can teleport ``an
unknown state of a quantum system which lives in a 2-dimensional Hilbert space"
(for concreteness, states of a spin 1/2 particle).  In Eq.  (1) however, instead
of states of a spin 1/2 particle we wrote ``qubit", where by qubit we understand
the quantum information which can be encoded in one spin 1/2 particle.  This
information need not be originally encoded in one spin 1/2 particle.  It could,
for example, be distributed among many spins.  Indeed, as Schumacher
and others showed~\cite{sch,schjoz,barnum}
quantum information can be efficiently manipulated --- compressed or
diluted essentially
without losses, similarly to classical information.  Thus it makes sense
to talk about ``the quantum information which could be compressed into
a spin
1/2 particle".

The question is whether we could replace the left-hand side of Eq. (1) by a 
similar abstract quantity. That is, we would like to be able to say something 
like

\begin{equation}
1~{\rm ebit}~teleports~1~{\rm qubit},
\end{equation} 
where 1 ebit describes any quantum system which contains entanglement equivalent 
to that of a singlet.

As a matter of fact, Eq. (2) is in common use. The point is that, at least for 
pure states, there are efficient ways in which entanglement can be manipulated, 
and arbitrary states can be transformed ---
essentially without losses --- into singlets 
\cite{entan}. 
Indeed, suppose that two distant observers, Alice and Bob, initially share a 
large number $n$ of pairs of particles, each pair in the same arbitrary state 
$\Psi$. Then, by performing suitable local operations and by communicating 
classically to each other, Alice and Bob can obtain from these $n$ copies of the 
state $\Psi$ some number $k$ of pairs, each pair in a singlet state. The action 
is ``essentially without losses" since Alice and Bob can
transform the $k$ singlets back  
into $n$ $\Psi$s. (The actions are reversible in the asymptotic limit of large 
$n$; the requirement of the asymptotic limit for reversibility is similar to 
that in compressing classical and quantum information.) The quantity of 
entanglement of an arbitrary state, measured in ebits, is simply $k/n$, the 
number of singlets which can be obtained reversibly from each pair of 
particles in the original state $\Psi$ \cite{entan,vonneumann}.

However, an important element is missing. While during concentrating and 
diluting entanglement by the efficient methods described in~\cite{entan},
entanglement is 
not lost, Alice and Bob might have to communicate classically to each other. 
They have thus to pay the price of exchanging some bits of classical 
communication.

The classical communication cost of entanglement manipulation is a largely 
ignored problem. Indeed, the general attitude is that entanglement is 
``expensive" while classical communication is ``cheap", and all the effort is 
generally directed only to preserving entanglement by all means. However,
to claim that entanglement is truly a fungible resource, one
must also consider the classical communication cost of entanglement 
manipulation.

The classical communication cost of entanglement manipulation has in fact 
implications for teleportation.  Indeed, Eq. (1) which describes the original 
teleportation is rather incomplete. The complete statement is that 

\begin{equation}
1~{\rm singlet}~+~{\rm communicating}~2~{\rm classical~bits}~teleports~1~{\rm 
qubit}.
\end{equation} 
Obviously, the more abstract equivalent of this equation, namely
\begin{equation}
1~{\rm ebit}~+~{\rm communicating}~2~{\rm classical~bits}~teleports~1~{\rm 
qubit},
\end{equation} 
or, following Bennett's notation, 
\begin{equation}
1~{\rm ebit}~+~2~{\rm bits} \geq 1~{\rm 
qubit},
\end{equation} 
would not be valid if during transforming the original supply of entanglement 
(in some arbitrary form) into singlets required for 
teleportation, Alice and Bob had to exchange supplementary bits of 
classical information. 

For teleportation the matter seems to be rather academic.
It is the entanglement which has the fundamental role, while the 
classical bits are, to a large extent, secondary --- in the absence of 
entanglement, no matter how many classical 
bits Alice and Bob exchange,
teleportation would be impossible. However, for other quantum communication 
tasks, the classical communication cost is highly relevant. Consider for example
the ``dense coding" communication method~\cite{dense}.
As Bennett and Wiesner 
showed,  when Alice and Bob share a singlet, Alice can communicate to Bob
two
classical bits by sending a single qubit. The basic equation is thus

\begin{equation}
1~{\rm singlet}~+{\rm communicating}~1~{\rm qubit}~communicates~2~{\rm 
classical~bits}, 
\end{equation}
whose mathematical abstraction is

\begin{equation}
1~{\rm ebit}~+~1~{\rm qubit} \geq 2~{\rm 
bits}. 
\end{equation}
In dense coding the main goal is to enhance the ability of performing classical 
communication  by using entanglement. However, if in the process of  
transforming the original supply of arbitrary entanglement into singlet form  we 
had to use a lot of classical communication, this would defeat the
objective of 
the entire exercise. 

In the present paper we show that for bi-partite pure states,
(in the asymptotic limit)
entanglement can be transformed --- concentrated and diluted --- in a reversible
manner with {\it zero classical communication cost}.  Hence, the notion of
``ebit" is completely justified.  In other words, it doesn't matter in which
form entanglement is supplied; all that matters is the total quantity of
entanglement.  Provided that they have the same von Neumann
entropy, both singlets and partially entangled states have the same power to
achieve any task in quantum information processing (in the asymptotic limit).

In order to establish entanglement as a fungible resource, we have to show that
both entanglement concentration (transforming arbitrary states into singlets)
and entanglement dilution (transforming singlets into arbitrary states) can be
done without any classical communication cost.  The first task is easy --- the
original entanglement concentration method presented in~\cite{entan}
proceeds without any
classical communication between the parties.  In other words,
the classical communication
cost of the procedure is identically equal to zero.  The rest of this paper
is devoted to studying entanglement dilution. We will show that,
although diluting entanglement may require classical communication, the
amount of communication can be made to vanish in the asymptotic limit.

The standard entanglement dilution scheme~\cite{entan}
requires a significant
amount of classical communication (two classical bits
per ebit). Therefore, it fails to demonstrate
the complete inter-convertibility of entanglement.
To establish entanglement as a truly fungible resource,
we present a new entanglement 
dilution scheme which conserves entanglement and requires 
an asymptotically vanishing amount of classical communication.
To construct our scheme, we first prove the following.

{\bf Lemma}: Suppose Alice and Bob
share $n$ singlets. Let  $\Pi$ be the state of a bi-partite system AB where each 
system has a 
$2^n$ dimensional Hilbert space, and let the Schmidt
coefficients~\cite{schmidt} of 
$\Pi$ be
$2^r$-fold degenerate.   Then, there is a procedure by which
Alice and Bob can prepare $\Pi$ shared between
them such that only $2 (n-r)$ bits of
classical communication and local operations are needed.

{\bf Proof}: With the $2^r$-fold degeneracy in Schmidt coefficients,
$\Pi$ can be factorized into
a direct product of $r$ singlets and a residual state, $\Gamma$, whose Schmidt 
decomposition contains only
$2^{n-r}$  terms, i.e., up to bi-local unitary transformations,
\begin{equation}
\Pi =  \Phi^r \otimes \Gamma ,
\end{equation}
where $ \Phi$ denotes a singlet state.
Since Alice and Bob initially share singlets $\Phi$, there is no
need to teleport the $\Phi$s.
To share $\Pi$ non-locally, Alice only needs to teleport
the subsystem $\Gamma$ to Bob.
Alice and Bob can then apply {\it bi-local} unitary
transformations to their
state to recover $\Pi$. (We do not know if such local computations
can be done efficiently, but this is unimportant here.)
Since the dimension of $\Gamma$ is only $2^{n-r}$,
only $2(n-r) $ bits are needed for its teleportation.

{\it Remark.} Compared with a direct teleportation of the whole state
$\Pi$, the above procedure provides
a saving of $2r$ classical bits of communication because of the $2^r$-fold 
degeneracy
of Schmidt coefficients.

The crux of this Letter is the following theorem.

{\bf Theorem}: In the large
$N$ limit, $N$ copies of any pure bi-partite state $\psi$
can be approximated with a fidelity~\cite{jozsa} arbitrarily
close to $1$ by a state
that has $D=2^d= 2^{[ NS - O (\sqrt{N})]}$
degeneracies in its Schmidt decomposition
where $S$ is the von Neumann entropy of a subsystem of $\psi$.
In other words, given any $\epsilon > 0$, for a sufficiently large
$N$, we have

\begin{equation}
\psi^N =  \Phi^d \otimes \Delta + u_2
\end{equation}
where $d= {[ NS - O (\sqrt{N})]} $,
$\Delta$ is an un-normalized residual state whose Schmidt decomposition contains 
 $2^{O(\sqrt{N})}$ terms,
and $\| u_2  \| < \epsilon $.

{\it Remark}: When combined with the Lemma, the Theorem
implies that Alice and Bob can perform entanglement dilution
from $N$ copies of $\psi$ to $NS$ singlets
using an asymptotically vanishing number, namely
$O (\sqrt{N}/N) = O ( { 1 / \sqrt{N} })$
of classical bits of communication per ebit.
This establishes the main result of this Letter.

{\bf Proof of the Theorem}: The idea of the proof is simple.
We would like to decompose the state $\psi^N$ into two pieces, $ \psi^N =
u_1 + u_2 $ such that
the dominant piece, $u_1$ has a large degree of degeneracy
in its Schmidt coefficients as required in the Theorem, while $u_2$ is small.

While the idea of our proof is general, it is best
understood by considering the special case when $\psi = a | 00 \rangle
+ b | 11 \rangle$.
Consider the Schmidt coefficients of $\psi^N$.
They have the form $a^k b^{N-k}$ 
and are, in general, highly degenerate --- the coefficient
$a^k b^{N-k}$ appears $ N \choose k$ times.

The first step of our proof is to note that we can divide the different values
of $k$ into two classes --- ``typical" and ``atypical". 
For a ``typical'' value of $k$,  $\log {N \choose k}$ lies between
$N S( \psi) - O ( \sqrt{N})$ and $N S( \psi) + O ( \sqrt{N})$,
say between $N S( \psi) - 10  \sqrt{N}$ and $N S( \psi) + 10  \sqrt{N}$.
(The actual coefficient of the $ \sqrt{N}$ term will depend on the
value of $\epsilon$
used in the Theorem. Here, we simply take it to be $10$ to illustrate
the basic idea of the proof.)
All other values of $k$ are ``atypical". It is well-known that,
compared to the measure of the 
typical set, the overall 
measure of the atypical set is very small.
(i.e. the norm of the projection of $\psi^N$ on the Hilbert subspace 
spanned by the atypical terms in the Schmidt decomposition is small).  We shall 
include all the atypical terms in $u_2$.

Let us now concentrate on the typical terms.  According to the requirement of
the theorem, all terms in $u_1=\Phi^d \otimes \Delta$ are degenerate and their
degeneracies have a common factor of the order of $2^d=2^{[ NS - O (\sqrt{N})]} 
$.
If the degrees of degeneracy of the typical terms all had a common factor
of the order of $2^{[ NS - O (\sqrt{N})]} $, we could include all these terms in
$u_1$, and the proof would be complete.  Unfortunately, although indeed each
term in the typical set has a degeneracy of the order $2^{[ NS - O (\sqrt{N})]} 
$,
when one varies $k$ over the typical set, the various values of $N \choose k $
do not have a large common factor. To deal with this problem we
``coarse-grain'' the number of terms of
Schmidt decomposition grouping them in bins of
say $2^{ \lceil N S( \psi) - 20  \sqrt{N}\rceil }$.
More concretely, for each $k$ in the typical set, let
the number of full bins $n_k$ be such that
\begin{equation}
 n_k 2^{ \lceil N S( \psi) - 20  \sqrt{N}\rceil} \leq {N \choose k} <
(n_k + 1) 2^{ \lceil N S( \psi) - 20  \sqrt{N} \rceil}.
\end{equation}
We simply keep
only $n_k 2^{ \lceil N S( \psi) - 20  \sqrt{N}\rceil}$
out of the original  ${N \choose k}$ terms
in $u_1$
and put the remaining ${N \choose k} - n_k
2^{ \lceil N S( \psi) - 20  \sqrt{N}\rceil} < 
 2^{ \lceil N S( \psi) - 20  \sqrt{N}\rceil }$ terms in $u_2$.
Now $n_k $ is at least of the order $2^{10  \sqrt{N}}$
and is, therefore, very large.
Consider $u_1$. The degeneracies of its Schmidt coefficients are
multiples of $2^{ \lceil N S( \psi) - 20  \sqrt{N}\rceil}$, hence we can write
 $u_1 = \Phi^d \otimes \Delta$
where $d =  \lceil N S( \psi) - 20  \sqrt{N}\rceil$.

Let us now summarize. By construction, the state $u_1$
is of the form $\Phi^d 
\otimes \Delta$. 
The norm $\| u_2 \|$ is very small for two reasons: 1) the
contribution to $\| u_2 \|$ from the atypical set is 
small and 2) for each $k$ in the typical set, its contribution 
to $\| u_1 \|$ is at least $n_k$ times its contribution to
$\| u_2 \|$ where $n_k$ is very large.
Consequently, $\phi^N = u_1+ u_2 = \Phi^d \otimes \Delta + u_2$
where $d= {[ NS - O (\sqrt{N})]} $,
$\Delta$ is an un-normalized residual state of $2^{O(\sqrt{N})}$ dimensions,
and $\| u_2  \| $ is very small.
\hfill Q.E.D.

In conclusion, we have shown that entanglement dilution from
$N[S ( \psi) + \delta]$ singlets to $N$ pairs of a bi-partite
pure state $\psi$ can be
done with only $O ( \sqrt{N})$ bits of classical communication.
So the number of classical bit per ebit needed
is $O ( { 1 \over \sqrt{N}})$, which vanishes asymptotically.
In other words, states with the same amount of bi-partite entanglement
are inter-convertible to one another in the asymptotic limit
(with vanishing amount of classical bits of communication per ebit).
Therefore, entanglement bits or ``ebits''
can be regarded as a universal quantum
resource, as originally proposed by Bennett and others.

The above discussion has been done for the case of pairs of two spin 1/2 
particles in pure 
states. The generalization to pure states of pairs of higher spin particles is 
immediate. However, generalization towards multi-particle entanglement and/or 
density matrices is problematic.

In the case of pure-state multi-party entanglement, not only do we not know
about the classical communication cost of transforming entangled states from a
form into another, but it is also not yet clear whether there exists a 
reversible
procedure which can transform (in asymptotical limit) $n$ copies of an arbitrary
multi-party pure state $\Psi$ into some standard entangled state (or set of
states \cite{multi}).  In fact, it is not even clear what the standard entangled
states should be.  The existence of such a procedure is, however, quite
probable.

The case of density matrices is even more complicated.  Here, even in the
simplest case of pairs of spin 1/2 particles, it is probable that reversible
transformations do not exist at all.  That is, although arbitrary entangled
density matrices can be prepared from singlets, and then singlets can be
reconstructed from the density matrices, the number $k_{in}$ of spins
necessary to create $n$ copies of an arbitrary density matrix is probably always
larger than the number $k_{out}$ of spins which can be obtained from the $n$
density matrices.  (Following the terminology of \cite{puri}, the entanglement
of formation is larger than the entanglement of distillation).  If indeed this
is the case, it is then probable that these transformations require 
non-negligible classical communication.  Actually, a reasonable
conjecture is that there exists a very close connection (possibly a sort of
conservation relation) between the amount of irreversibility in the
transformation {\it singlets} $\rightarrow$ {\it density matrices}
 $\rightarrow$ {\it singlets}
and the amount of classical communication needed for this process.

Finally, we would like to add some more general remarks. If we restrict the 
actions one is allowed to perform on the entangled states, entanglement might no 
longer be inter-convertible. For example, if we do not allow {\it collective} 
processing but insist that each pair of entangled particles should be processed 
separately, then entanglement is not inter-convertible anymore. Indeed, while 
one could still produce singlets from partially entangled states such as $\alpha 
|1 \rangle |1 \rangle +\beta|2 \rangle|2 \rangle$
by using the procustean method  \cite{entan}, this action 
is not reversible (that is, the overall probability of success for the chain of 
actions {\it initial state} $\rightarrow$ {\it singlet} $\rightarrow$ {\it 
initial state} is less than 1). 

Thus entanglement is a fungible resource only when no restrictions are placed on
the allowed entanglement manipulation procedures. This raises the question of 
what
exactly do we mean by the ``unrestricted" set of actions?  The usual paradigm
\cite{entan,puri,martin,thermo} of manipulating entanglement is that of 
``collective local
actions + classical communication", and the basic statement is that:

{\it ``Entanglement cannot increase by collective local actions and classical 
communications."}

However, in the light of the new effects discovered by R., P., and M.
Horodecki, that is, the existence of bound entanglement \cite{bound} and
especially the possibility of activating bound entanglement \cite{active} this
paradigm might turn out to be insufficient.  And, indeed, it is very
restrictive.  After all, why not allow also {\it quantum communication}?  It is
true that quantum communication does not conserve entanglement and permits
creation of entanglement out of nothing.  However, there is no reason why such
non-conservation could not be easily kept under control.  We would thus suggest
the paradigm of ``collective local actions + classical communication+ quantum
communication", and the basic statement that

{\it ``By local actions, classical communications and $N$ qubits of quantum 
communication, entanglement cannot increase by more than $N$ e-bits."}

{\bf Acknoledgements} We would like to thank C.  H.  Bennett, D.  Gottesman, and
R.  F.  Werner for helpful discussions.  Part of this work was completed during
the 1997 Elsag-Bailey -- I.S.I.  Foundation research meeting on quantum
computation.


\begin{references}
\bibitem{vonneumann} This number turns out to be nothing 
else than the von Neumann entropy of the reduced density matrix of (either of) 
the 
members of the pair of entangled particles.
\bibitem{schmidt} Let $H_A$ and $H_B$ be Hilbert spaces with
dimensions $p$ and $q$ respectively. Consider a normalized
state $| \Phi \rangle $ in $H_A \otimes H_B$.
Let $\rho = | \Phi \rangle \langle \Phi | $ be the density matrix and
$\rho^A = {\rm Tr}_B \rho$ and $\rho^B = {\rm Tr}_A \rho$ be the
reduced density matrices. Then the Schmidt decomposition
theorem \cite{JMO}
states that $| \Phi \rangle $ can be written as
\begin{equation}
| \Phi \rangle = \sum_{i =1}^r \sqrt{ \lambda_i } | a_i \rangle
\otimes | b_i \rangle ,
\end{equation}
where $| a_i \rangle$ ($| b_i \rangle$ respectively) are orthonormal
eigenvectors of $\rho^A$ ($\rho^B$ respectively), and $r < {\rm min}
(p, q)$ is the total dimension of the non-zero eigenspaces
of $\rho^A$. The coefficients $\sqrt{ \lambda_i }$ are called the
Schmidt coefficients.
\end{references}
\end{document}